%% file: POLW.TEX
\input A4M.tex

\hoffset=-0.5cm

\rulewidth0pt 
\def\e{{\rm e}}


\cl{\bfbb Statistical mechanics of non-hamiltonian systems:}

\vskip3pt
\cl{\bfbb Traffic flow}

\bigskip\bigskip
\cl{\rmb Anton \v Surda}
\bigskip

{\ninepoint
\cl{\it Institute of Physics, SAS, 815 00 Bratislava, Slovakia}

\bigskip\bigskip

\medskip
\hrule
\medskip

\ni{\bf Abstract:} 
Statistical mechanics of a small system of cars on a single-lane road is 
developed. 
The system is not characterized by a Hamiltonian but by a conditional 
probability 
of a velocity of a car for the given velocity and distance of the car ahead.   
Distribution of car velocities for various densities of a  group of cars 
are derived as well as probabilities of density fluctuations of the group for 
different velocities. For high braking abilities of cars free-flow and 
congested phases are found. Platoons of cars are 
formed for system of cars with inefficient brakes. A first order phase 
transition between free-flow and congested phase is suggested. 
 
\bigskip
\ni{\it Keywords:  non-Hamiltonian systems, microcanonical, canonical, traffic 
flow}

\bigskip

\ni{\it PACS:  05.20.Gg, 05.50.+g,  05.60.Cd,  89.40.Bb} 

\vfootnote{*}{E-mail: anton.surda@savba.sk}

}

\bigskip

\hrule\hrule\hrule

\bigskip\bigskip

\ni{\bfb1 Introduction}
\bigskip

\rm
\ni In classical equilibrium statistical mechanics 
Hamiltonian  plays a decisive role in 
determination of  statistical properties
of a system of interacting particles. 
In most cases it represents  energy 
as a function of momenta and spatial coordinates of the particles.  
The system may be treated microcanonically or canonically. 
In 
a microcanonical ensemble all the states are equally probable and the number 
of states  of a given energy determines the entropy of the system. In 
a canonical ensemble  the probability of states of a subsystem  are 
proportional to $\e^{-\beta E_{\rm s}}$, where $\beta$ is the inverse 
temperature and $ E_{\rm s}$ is the energy of the subsystem. The sum of 
probabilities of all the states is called partition function and determines 
thermodynamic properties of the system. The subsystem may be taken as small as 
one particle, and the  expression $\e^{-\beta (E_i(q_i,r_i)+E(C_{i,j})} $ 
determines the conditional probability that a particle has momentum $q_i$, 
coordinate $r_i$ if its energy  is $E_i$ and interaction energy 
with  neighbouring particles of generalized coordinates $C_{i,j}$ is 
$E_{i,j}(C_{i,j})$. 

 In our approach, we do  not start from the Hamiltonian of a system of 
particles, which we is not even introduced, but from the knowledge of 
conditional probabilities of generalized coordinates of each particle.  It 
is assumed that the conditional probability depends only on the state of 
the neighbouring 
particles, and it can be determined from the behaviour of a small system 
experimentally. 

This non-Hamiltonian approach is  applied to a 
system of cars  on a single-lane road while the behaviour of the cars is 
described 
by conditional probabilities of the velocity of each car depending on the 
distance and velocity of the car ahead.

Traffic flow  of a system of identical cars on a single-lane road
has been intensively studied in the recent decade using dynamical or 
kinetic 
description of car behaviour [1,2].  The models used were continuous (fluid 
dynamical 
models), car-following models [3],  or discrete particle hopping models 
related to cellular 
automaton models with  stochastic behaviour [4].

In the approach of Mahnke et al. [5]  the group of cars is represented by a 
grandcanonical 
ensemble,  number of cars in which is not fixed, and its chemical potential 
is a function of parameters of a master equation.

In the last years we could observe  a revival of the microcanonical approach to
the problems of statistical mechanics [6--10].  One of the reasons for that 
was identification
of the 
region where the entropy of a finite system is convex, instead of the standard 
concave 
shape of it,  with a point or line in the phase space where the 
first-order phase transition in the corresponding infinite system takes place. 
 As the number of observed cars in normal traffic is not too large, the 
techniques developed in statistical physics for small systems are convenient 
in this case. The term ``phase transition'' in this paper is used in the sense 
of the above-cited works.

Statistical mechanics of 1D non-Hamiltonian models is developed in Section 2. 
In Section 3 a traffic model  inspired by the above-mentioned 
car following 
 and discrete particle hopping models is introduced. In Sections 3 and 4
formulas for microcanonical and canonical treatment of our model are derived. 
The results for a subsystem of 5 cars are presented  in Section 5. 

\bigskip\bigskip

\ni{\bfb 2 Statistical mechanics of 1 D non-Hamiltonian systems}

\bigskip
\ni A single-lane road represents a one-dimensional system so we 
shall further confine ourselves to 1D systems of particles with nearest 
neighbour 
interactions. The conditional probability that the particle has velocity $v_i$ 
and coordinate $r_i$ while the velocity and coordinate of its neighbouring 
particles are $v_{i\pm 1}$ and $r_{i\pm 1}$, respectively, will be denoted as 
$p(v_i,r_i|v_{i-1},r_{i-1}, v_{i+1},r_{i+1}) \equiv 
p(v_{i-1},r_{i-1},v_i,r_i,v_{i+1},r_{i+1}) 
/ p(v_{i-1},r_{i-1},v_{i+1},r_{i+1})$ and taken as an input to the theory.  
The probability of  velocities and coordinates of all  $n$ particles of the 
system   can be easily calculated from slightly different conditional 
probabilities 
$$
p(v_1, r_1,\dots,v_n, r_n)= p(v_1,r_1|v_2,r_2)p(v_2,r_2|v_3,r_3)) 
 \dots  
p(v_{n-1},r_{n-1} |v_n,r_n)p(v_n,r_n)  \eqno(1)  
$$
where $p(v_i,r_i|v_{i+1},r_{i+1}) = p(v_i,r_i,v_{i+1},r_{i+1}) / 
p(v_{i+1},r_{i+1})$. 

Probabilities
$p(v_i,r_i|v_{i+1},r_{i+1})$ are unknown, but they are closely related to the 
probabilities $p(v_i,r_i|v_{i-1},r_{i-1}, v_{i+1},r_{i+1})$
characterizing our system. They can be calculated from the following 
system of equations:  
$$
\eqln{
& p(v_i,r_i|v_{i+1},r_{i+1})=\sum_{v_{i-1},r_{i-1}}
p(v_{i-1},r_{i-1}|v_{i+1},r_{i+1})p(v_i,r_i|v_{i-1},r_{i-1},v_{i+1},r_{i+1}) 
\cr &p(v_{i-1},r_{i-1}|v_{i+1},r_{i+1})=\sum_{v_i,r_i} 
p(v_{i-1},r_{i-1}|v_{i},r_{i})p(v_i,r_i|v_{i+1},r_{i+1}).& (2) \cr}
$$

\ni $p(v,r)$ in a homogeneous system can be obtained by summation of 
$p(v_1, r_1,\dots,v_n, 
r_n)$ over all coordinates of the system but one. In a finite system 
$p(v_n,r_n)$ 
 represents a boundary condition,  which in our case is
the probability  of the velocity and coordinate of the first 
particle of a large reservoir. The spatial coordinates of the reservoir 
particles  satisfy $r_m>r_{n-1}$. 

In the system of cars on a single-lane road  the driver behaviour  is assumed 
to 
depend only on the previous car. In such a system the conditional probability 
of velocity and coordinate of particle 1 depends only on the  velocity and 
coordinate of particle 2 ahead of it: 
  $p\equiv p(v_1,r_1|v_2,r_2)$.
Then the equations (3) are superfluous and the probabilities of the whole 
system are given directly by (2). 
These probabilities will be used for calculations of the sum of probabilities 
of the system as a function of length and  total velocity (sum of 
velocities of all cars of the system).

The terms microcanonical and canonical will be used  
in the case of non-Hamiltonian systems as well, but they will be related  
rather to the {\it length} of the system than
to  its {\it energy}, which is not introduced.

In the canonical ensemble of a Hamiltonian system the usually calculated 
quantity is the partition 
function, which is sum of the probabilities of all possible states of the 
system. In the microcanonical ensemble, the probability to find the system in 
a state is same for all of them. Logarithm of their sum for some 
fixed 
quantities is called entropy. In our non-Hamiltonian approach, we calculate 
sum of the probabilities of states (SPS) as a function of length of the 
system and  sum  of velocities of the particles.  

In the microcanonical approach the length of a group of cars is fixed, and 
it is influenced  by the reservoir  only through boundary conditions. 
The properties of the system are given by the sum of the probabilities of 
states with the same total velocity.

In the canonical approach the group of cars represents only a subsystem of a 
large system whose remaining part is a reservoir. The length of the whole 
system is constant, and the length of the subsystem is changed only at the 
expense of the length of the reservoir. The properties of the subsystem are 
given by sum of the probabilities of states  of the {\it whole system} having 
the same  total velocity of the cars in the {\it subsystem.} 
The canonical approach is able to describe fluctuations of the length of the 
subsystem around its mean value.  
 
\bigskip\bigskip

\ni{\bfb 3 Model} 

\bigskip
\ni The cars are further represented by dimensionless points moving on a 
discrete 
one-dimen\-sio\-nal lattice and are characterized by 2 quantities:  discrete 
velocity $v_i$ in the interval $\langle0, v_{\rm max}\rangle$ and a discrete 
coordinate (site number)   $x_{i} \in 
\langle1, X-1\rangle$. 
$v_{\rm max}$ is the maximum  velocity given by the construction of the car,  
and $X$ is the length 
of the observed group (subsystem)  of cars. The coordinate of each car is 
measured with respect to the last car of the group. Its coordinate is 
always 0, i.e.,  the origin of the coordinate system is moving with it. As the 
length of the group is $X$, the coordinate of the 
last car of the group ahead is also $X$. Number of cars in  the group is $N$. 
(The lattice constant is related to the car length).
Car velocities and coordinates acquire only integer values.

Kinetics of the system of cars is given by reaction of each  driver on the 
 car ahead moving with velocity $v_{i+1}$ at the distance $x_{i,i+1}$. 
As the driver directly influences only velocity of his car
his reaction is characterized by a conditional probability of a car velocity 
$v_i$ parametrized by velocity and distance of the car ahead:  $p\equiv 
p(v_i| x_{i,i+1},v_{i+1})$. The probability could be found experimentally by 
long observation of two cars at all possible situations. Since such data are 
not available yet, the probability is calculated from a simple model 
behaviour of a driver.

The  probability distribution is assumed to be
peaked around an optimal velocity  $v_{\rm opt}$,
which  is further chosen as 90\% of maximal safe velocity $v_{\rm m}$.
The maximal safe velocity is determined from the requirement that two 
neighbouring cars, which start to decelerate at the same time with the  same 
deceleration rate $a$, would stop without crash. Moreover,   $v_{\rm m}$ must 
not be greater than the maximum possible velocity of the car  $v_{\rm max}$, 
i.e., for every car 
$$
\eqln
{& v_{\rm opt}(v_2,x_{1,2}) = 0.9 v_{\rm m}(v_2,x_{1,2}) ,\cr 
&  v_{\rm m}(v_2,x_{1,2})=\cases{ -a\tau +
\sqrt{(a\tau)^2+ 2ax_{1,2}+v_2^2}& {\rm if}\  $v_{\rm m}\le v_{\rm 
max}$\cr
v_{\rm max}\quad & {\rm if}\ $ v_{\rm m}> v_{\rm max}$\cr}& (3) \cr}  
$$
 where $x_{1,2}$ and $v_2$ are the distance (headway) and velocity of the car 
ahead, respectively, $\tau$ is the reaction time of the driver of the car 1.  
(Problem  of non-zero reaction time is discussed in more 
detail in [11].)    As we use only integer values of 
velocities, the nearest integer value to $v_{\rm opt}$ from (1)  is taken    
for the actual optimal velocity in our further calculations. For very high 
densities, when  discreteness of the lattice is not negligible,  
another condition  $v_1< v_2+x_{1,2}$ must be  imposed.

The way of driving of the observed drivers is characterized by 
 distribution of probabilities of car velocities around the optimal velocity.
Here we use an extremely simple distribution, in which the probability of 
optimal velocity is $p_0$, the probabilities of the velocities $ v_{\rm opt}\pm
1$ are $p_1$, while the probability of the car to have any other 
permitted velocity is $p_2$. The values of the probabilities for velocities 
higher then the maximal safe velocity are equal to 0. The sum of all 
probabilities for each 
car is equal to 1. The parameters $p_0, p_1$ and $p_2$ are the same for every
car, 
and the distribution depends on the headway only by means of the value of 
optimal velocity. 

\bigskip\bigskip

\ni{\bfb 3 Microcanonical description}

\bigskip
\ni In the microcanonical approach  only such groups of $N$ cars, which 
length is $X$ and sum of their velocities is $V$, are studied. These groups of 
cars are not totally isolated. They
are influenced by  the velocity distribution of the car ahead of them with 
spatial 
coordinate $X$. The probability distribution of each car is given by the rule
above as a function of headway and the velocity of the car ahead, while the 
distances between them are arbitrary and limited only by the length of 
the group.

The sum of probabilities of states (SPS)  with total velocity $V$ of a group 
of $N$ cars and 
length $X$ will be denoted as $W(V,X)$. 
It can be calculated, using (2),  recurrently 
$$
\eqln{
& W_1(V_1, X_1 v_2) =p(V_1| X_1,v_2)\cr
&  \vdots\cr
&  W_i(V_i, X_i; v_{i+1}) =\sum_{v_i, x_{i,i+1}}
W_{i-1}(V_i-v_i, X_i-x_{i,i+1}; v_i) p(v_i| x_{i,i+1},v_{i+1})\cr
&  \hbox{for } i=2,N-1\cr
&  \vdots&  (4)\cr 
&  W_N(V, X; v_{N+1}) =\cr
& \qquad {} =\sum_{v_N,  x_{N,N+1}}
W_{N-1}(V-v_N, X-x_{N,N+1}; v_N) p(v_N| x_{N,N+1},v_{N+1})\cr
 & W(V, X) =\sum_{v_{N+1}} W_{N}(V, X; v_{N+1}) p(v_{N+1})\cr
}
$$
where $0\le v_j \le v_{\rm max}$, $0\le V_j \le j v_{\rm max}$,  
$j\le x_{j,j+1}, X_j\le X - j$. Probability $p(v_{N+1})$ in the last line of 
(4)  is the 
velocity probability of the last car of a large group ahead (reservoir)  of 
the studied 
group   with the same car density. 

SPS  in the reservoir of length $L_{\rm r}$, 
number of 
cars $N_{\rm r}$, with the density $\dst{ N_{\rm r}\over L_{\rm r}}= {N\over 
X}$, and fixed velocity of the last car $v_{N+1}$ is   
$$
W_{\rm r}(v_{N+1},L_{\rm r})   =  \sum_{v_{N+2},\dots,v_{N+N_{\rm r}}\atop
x_{N+1,N+2},\dots,x_{N+N_{\rm r},N+N_{\rm r}+1}}              
\prod_{i=N+1}^{N+N_{\rm r}}   p_i(v_i| x_{i,i+1}, v_{i+1})\delta_{L_{\rm 
r},\sum x_{i,i+1}}  \eqno(5) 
$$
It depends, in principle, on the velocity of the first car of the reservoir, 
but  numerical calculations show that for large $N_{\rm r}$ this dependence 
is
negligible. The probability $p(v_{N+1})$ is the normalized  SPS
$$
p(v_{N+1})= W_{\rm r}(v_{N+1},L_{\rm r})/\sum_{v_{N+1}}W_{\rm 
r}(v_{N+1},L_{\rm r})                                           \eqno(6) 
$$

The quantity $W(V,X) $ in (4)  comprises the probability that the sum 
of velocities of the 
cars in the group is $V$ and the number of possible configurations of 
occupation of $X$ sites by $N$ cars. It is, in fact,  a 
product of the probability and number of configurations $\Omega $
$$
W(V,X) ={W(V,X)\over\sum_V W(V,X)}\cdot \sum_V W(V,X) \equiv P(V,X)\cdot  
\Omega(X).                                                           \eqno(7) 
$$
Since for the fixed length of the subsystem, $\Omega(X)$ is constant, only the 
normalized probability  $P(V,X)$ is presented in  Results.

In the microcanonical approach  only subsystems of cars with the constant 
density, the same  as is the mean density of the whole  system, are studied. 
To take into account 
also the density fluctuations, it is more convenient to use the canonical 
description with variable density of the subsystem due to its variable length.

\bigskip\bigskip
\ni{\bfb 4 Canonical description}

\bigskip
\ni In canonical approach the length of the subsystem varies,
 only the length of the whole system, subsystem +  
reservoir  is fixed. The number of cars in the subsystem and in the reservoir 
remains constant, so the density of cars varies with varying length of the 
groups. 
Our canonical description differs from the grandcanonical approach of Mahnke et
al.
[5]  where  the density of the 
subsystem changes due to exchange of cars between the subsystem and reservoir.

In statistical mechanics the properties of a  reservoir are usually not 
calculated, 
only the values of derivatives of its entropy (logarithm of number of states)  
with respect to the quantities, which are fixed in the whole system, are  
assumed to be known. They are, e.g., temperature, chemical potential, etc. 
Similarly, in our
 canonical description of the  system of cars, a pressure of  reservoir 
exerted on the 
subsystem could be introduced. Nevertheless, this quantity cannot be directly 
measured, and it would depend on the velocity of the last car of the 
reservoir, so we prefer a direct calculation of number of states of a large 
enough reservoir for given velocity of the last car and  length of the 
reservoir. 
 
The length of the system $L_{\rm s}$ is the sum of the length of the subsystem 
and 
reservoir $X+L_{\rm r}$. The number of cars in the  subsystem and 
reservoir  are 
fixed and  denoted as $N$ and $N_{\rm r}$, respectively. If $X\ltlt L_{\rm r}$
and $N\ltlt N_{\rm r}$, the properties of the subsystem does not depend on 
velocity of the first car in the reservoir. 
 
SPS of the reservoir at given velocity of the 
last car  is calculated according to (5). 
SPS of the whole system at given total velocity 
of the 
subsystem $V$ and its  length $X$ can be obtained by the same way  as in the 
microcanonical case, only in the last term in (3) -- the probability 
of the velocity of the last reservoir car $p(v_{n+1})$ -- is replaced by   
SPS
of the reservoir. Last line of (3) now reads
$$
W(V, X) =\sum_{v_{n+1}} W_{n}(V, X; v_{n+1}) W_{\rm r} (v_{n+1},L_{\rm s} - 
X). 
$$
The mean density of the subsystem $N/\langle X\rangle$ is equal to the density 
of the whole system $(N+N_{\rm r})/  L_{\rm s}$.

The main difference between the microcanonical and canonical treatment is that 
in the first case only number of states of the subsystem is calculated  while 
in the latter case the properties of the subsystem  are given by the number of 
states of the whole system. In the microcanonical approach the reservoir is 
used 
only for calculation of boundary condition -- probability distribution
of the last car of the  subsystem.

\topinsert
\Ps{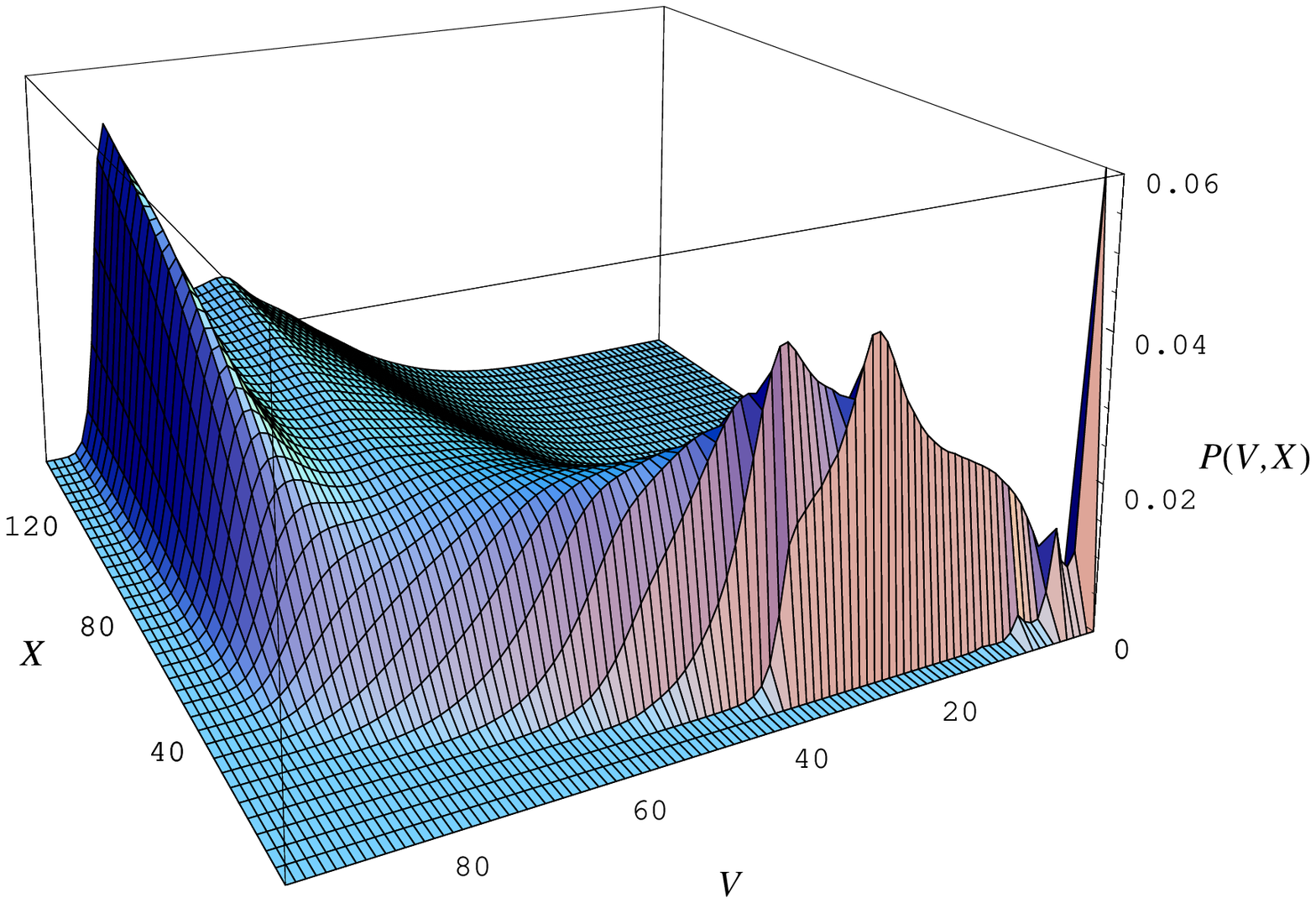}{0.6}
{Fig.~1. Probability of the total velocity  of a microcanonical ensemble 
of 5 cars as a function of its  total velocity $V$ and length $X$ for $a=4.0$, 
$p_2/p_0= 0.025$, $\tau=0$. The density of the system varies from 0.033 to 1.}

\topinsert
\Ps{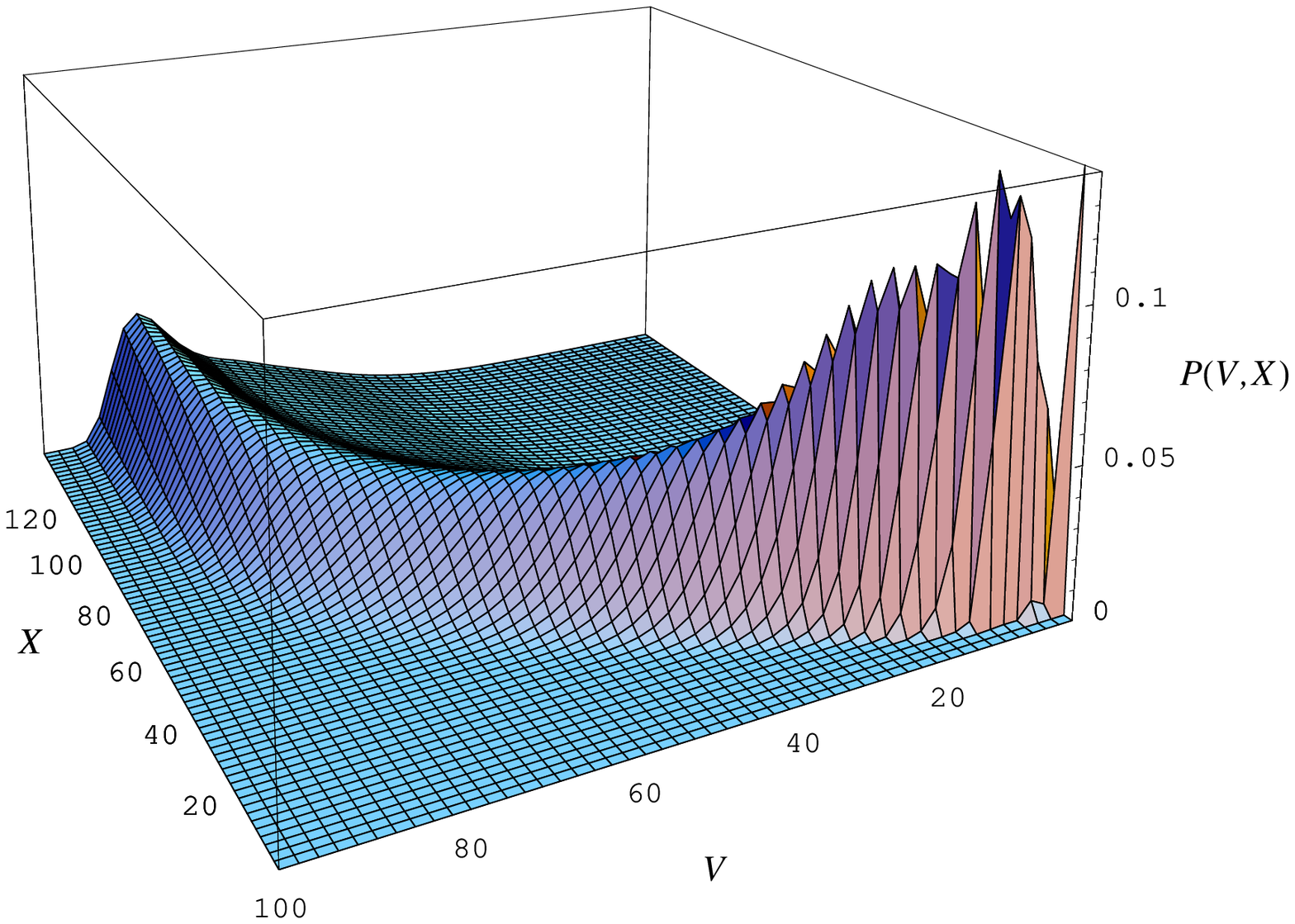}{0.6}
{Fig.~2. Probability of the total velocity  of a microcanonical ensemble 
of 5 cars as a function of its  total velocity $V$ and length $X$ for $a=4.0$, 
$p_2/p_0= 0.015$, $\tau=0.5$. The density of the system varies from 0.033 to 
1.}

\pageinsert
\Psn{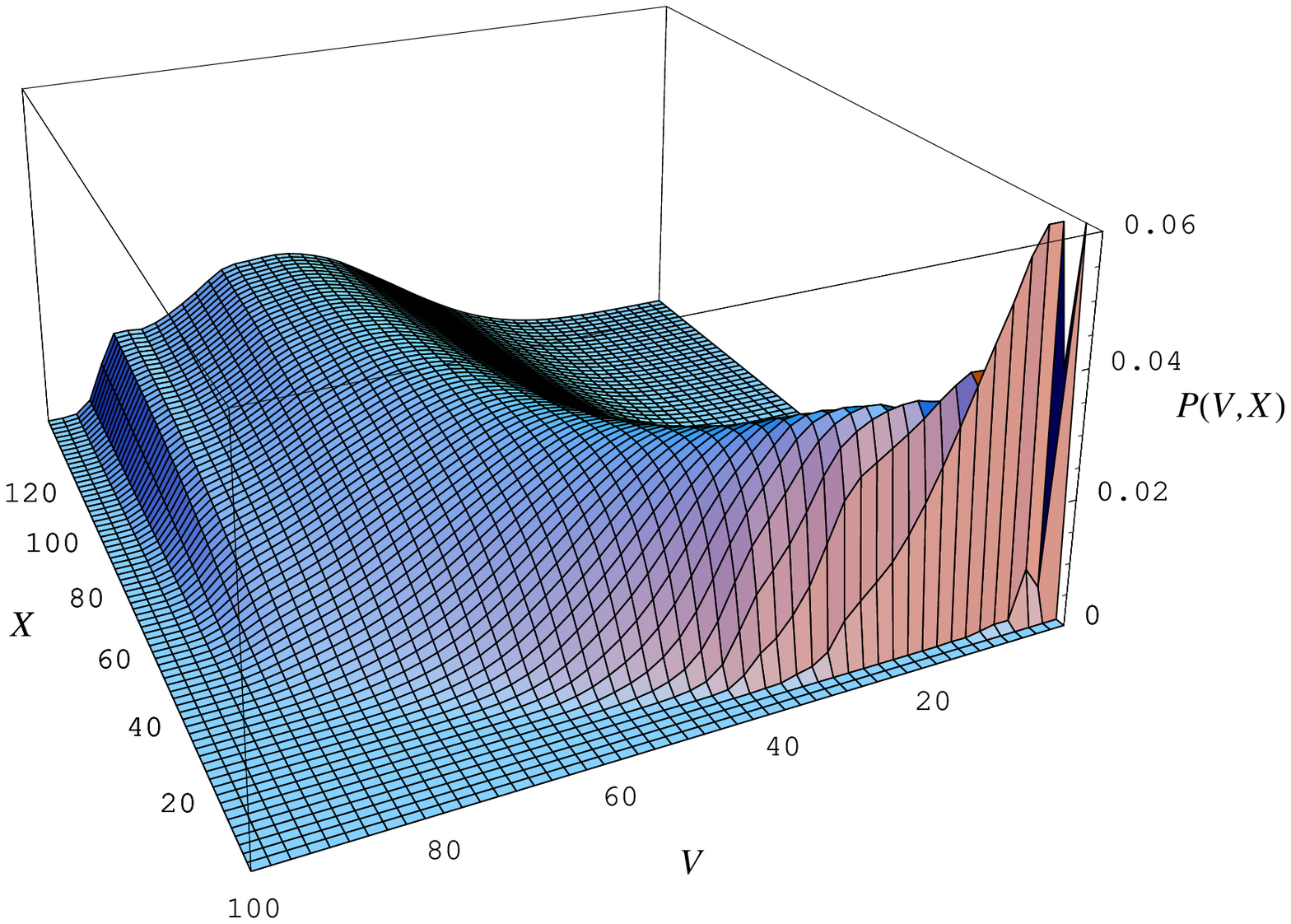}{0.6}
{Fig.~3. Probability of the total velocity  of a microcanonical ensemble 
of 5 cars as a function of its  total velocity $V$ and length $X$ for $a=4.0$, 
$p_2/p_0= 0.06$, $\tau=0$. The density of the system varies from 0.033 to 1.}

\Ps{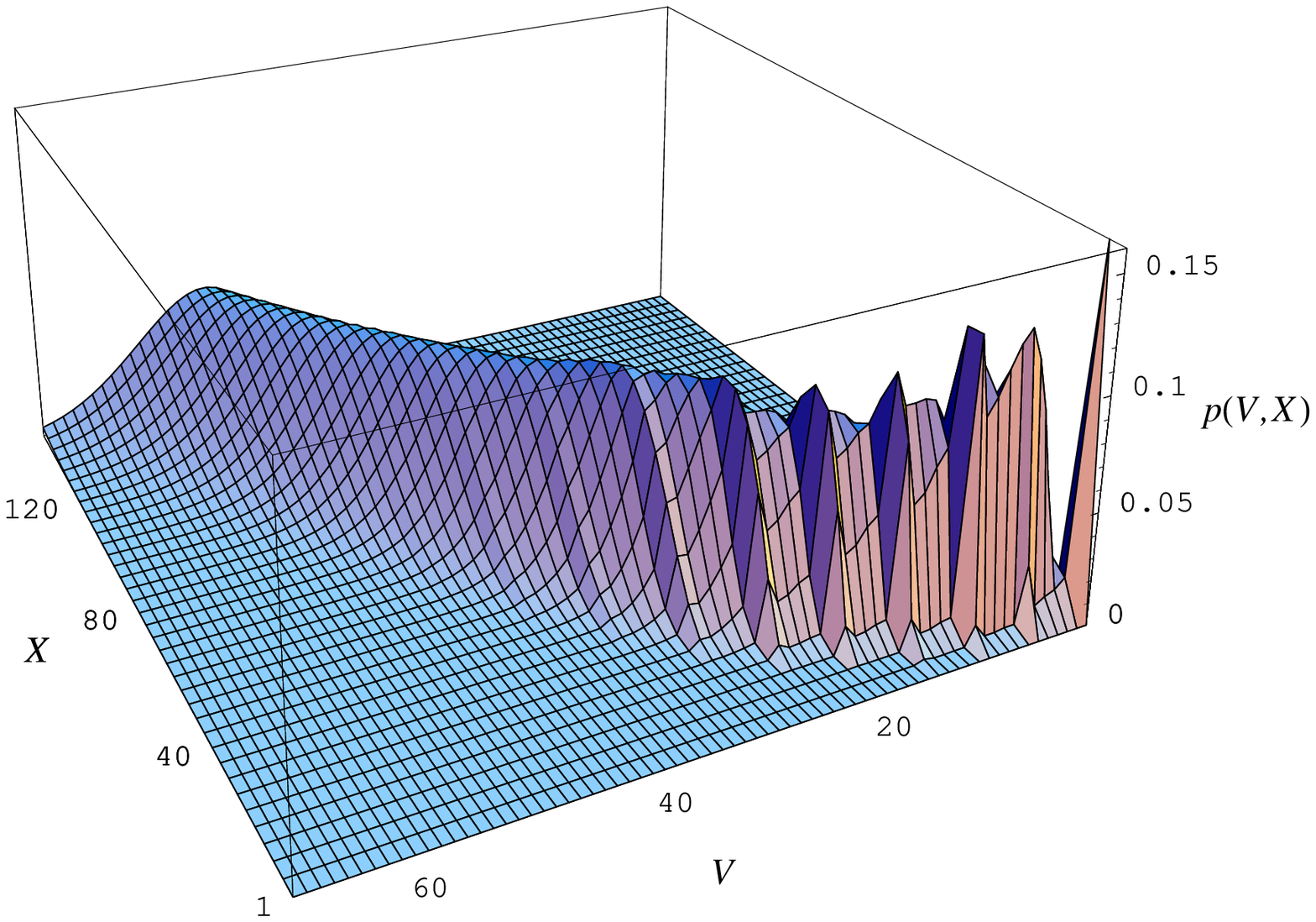}{0.6}
{Fig.~4. Probability of the total velocity  of a microcanonical ensemble 
of 5 cars as a function of its  total velocity $V$ and length $X$ for $a=0.5$, 
$p_2/p_0= 0.005$, $\tau=0$. The density of the system varies from 0.033 to 1.}

\topinsert
\Ps{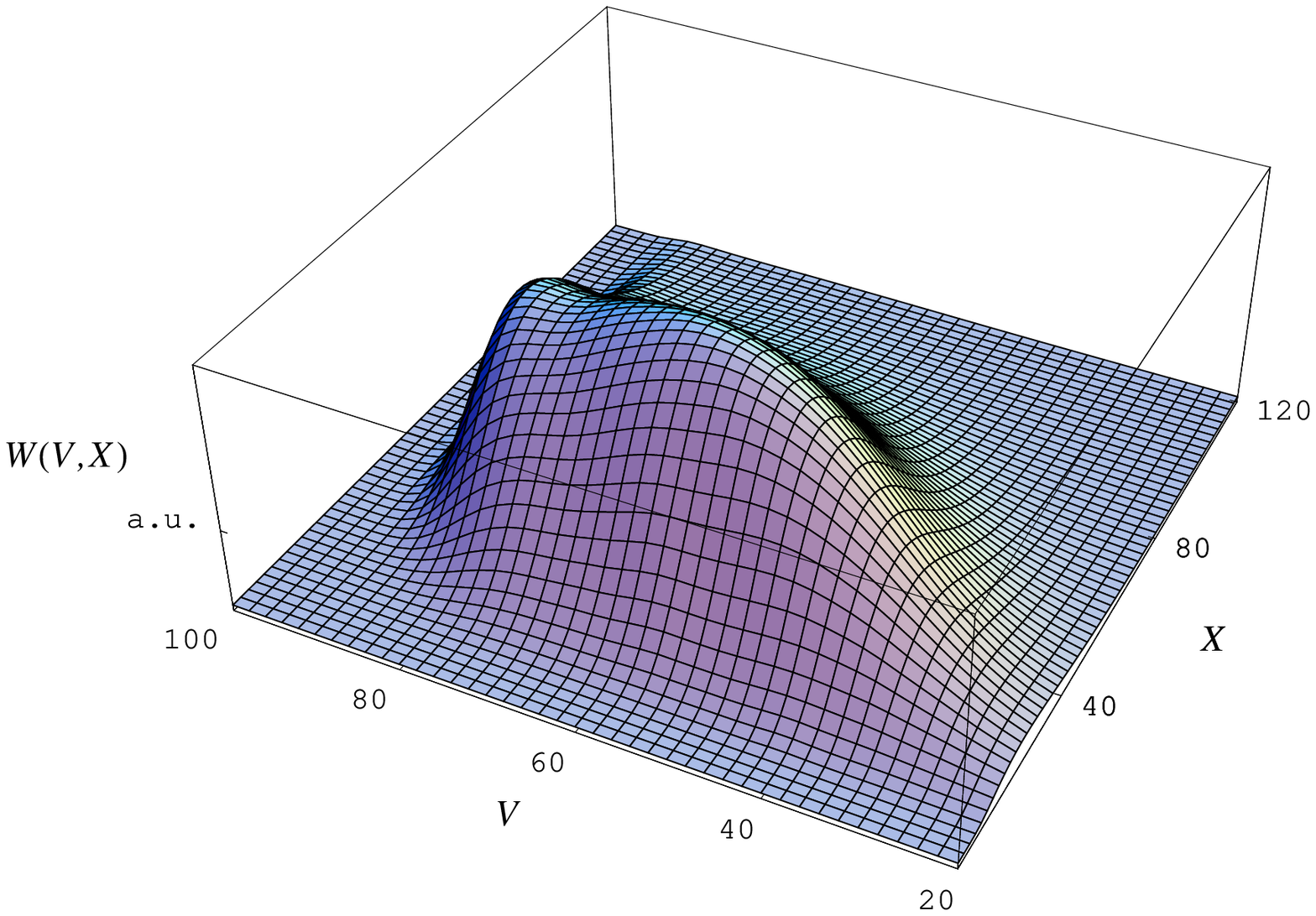}{0.6}
{Fig.~5. SPS of a canonical ensemble of 5 cars as a 
function of its 
total velocity $V$ and length $X$ for $a=4.0$, $p_2/p_0= 0.025$, $\tau=0$, and 
mean length $\langle X\rangle=50$, i.e., mean density $\rho=0.1$.}

\topinsert
\Ps{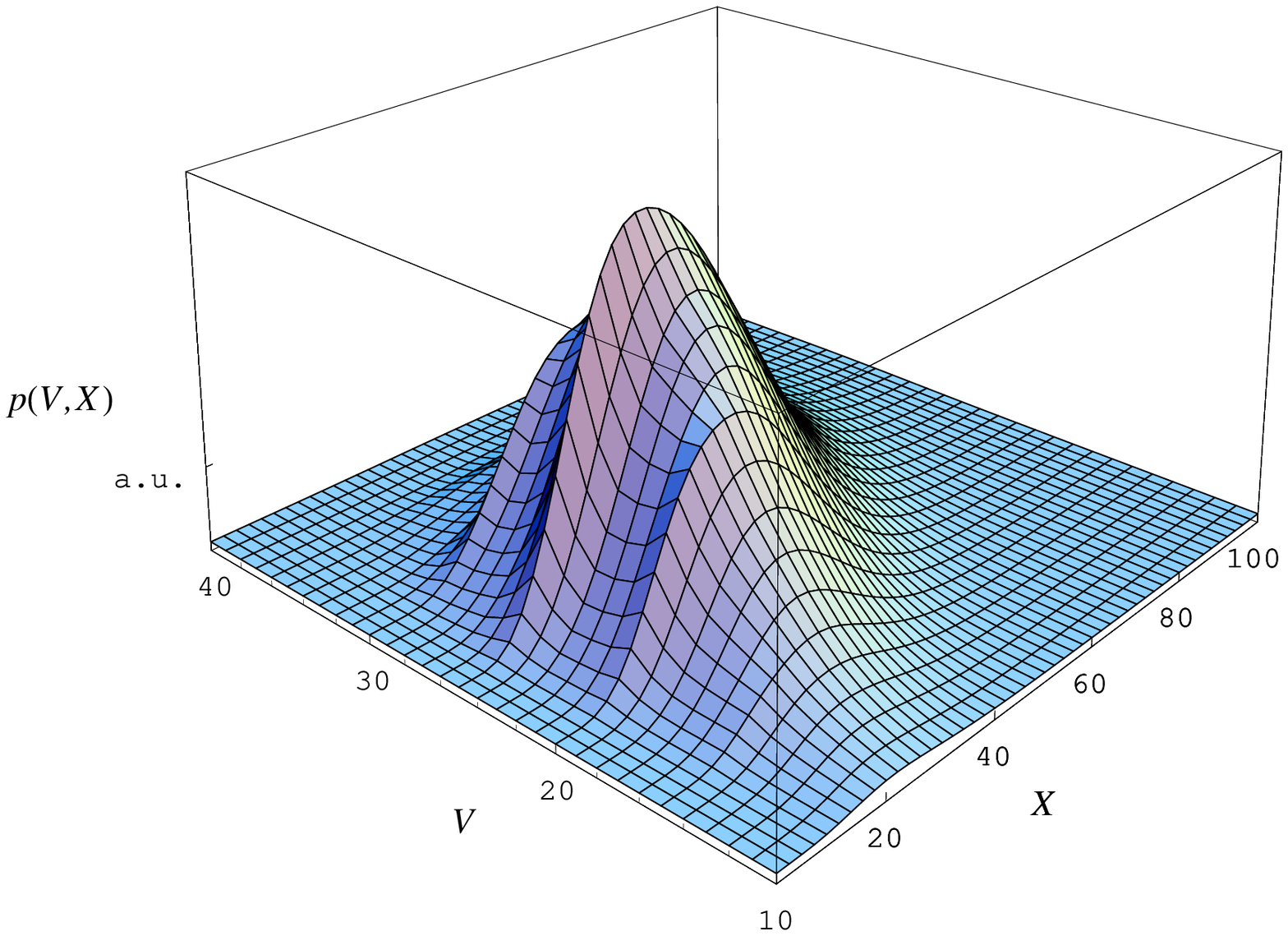}{0.6}
{Fig.~6. SPS of a canonical ensemble of 5 cars as a 
function of its 
total velocity $V$ and length $X$ for $a=0.5$, $p_2/p_0= 0.01$,  $\tau=0$, and 
mean length $\langle X\rangle=40$, i.e., mean density $\rho=0.125$.}

\bigskip\bigskip
 
\ni{\bfb 5 Results and discussion} 

\bigskip\ni
The velocities and positions of cars are described by discrete variables in our
model. Changing the values of its parameters, we can observe two 
different types of behaviour. In the first one,  for high deceleration 
rate $a$ and low densities, the system behaves like a continuous one;  in the 
SPS and probability of total velocity
 the underlying discrete structure of velocities is not seen. At 
small $a$ and high densities,  total velocities of the system, which  are 
integer multiples of number of particles, are more probable then the others.  
Platoons of cars of the same velocities are formed.
This regime reminds a ferromagnetic Potts model where the total 
magnetization of the system points in many different directions of the space.

In all our calculations, car velocity acquires 21 values $v_i=i,\ i=0,20$. The 
probability of a car to move with a velocity $v_i$ depends on the velocity and 
distance of the car ahead by means of optimal velocity $v_{\rm opt}(v,x)$.  
As stated above it acquires 3 values $p_0,p_1$ and $p_2$

In the present calculations $p_1/p_0$ is fixed to 0.3 and $p_2$ is chosen  
for 
the  free parameter, and the normalization 
condition is used. 

The position of a car 
with respect to the 
first one may be an integer between 1, and $X-1$ if the site is not occupied 
by another car. A subsystem of 5 cars as a part of a system of 50 cars is 
further studied.

The main result of our calculations is the probability of the total velocity 
$P(V,X)$ of the 
subsystem as a 
function of the subsystem length $X$ for fixed number of cars in the 
microcanonical ensemble and  
SPS $W(V,X)$ in 
the canonical case.
 They are plotted in 3D graphs.

In the microcanonical case the density of the subsystem is the same as 
the density of the reservoir. The probabilities of the sum of the velocities 
 of 5 cars   for lengths of the group from 5 to 150, i.e. for car densities 
from 1 to 0.033, 
 $a=4$, $p_2/p_0=0.025$, $\tau=0$ are shown in Fig.~1. This case represents a 
group of cars with good brakes (large deceleration rate $a$),  medium spread 
of velocities of each car, and zero reaction time of the drivers. For high 
group length (low density)  the cars mostly move freely  with optimal 
velocity, nevertheless due to the car velocity spread, there is a nonzero 
probability of smaller velocities. With increasing density, number of cars 
with small velocity  increases, and at $X=52$ the most probable velocity 
discontinuously drops to a value smaller by 20 units. In analogy with to 
microcanonical approach to finite systems of statistical mechanics where 
the entropy is not concave [6], we call 
this phenomenon first order phase transition. For length of the group $<30$, 
the total velocity decreases fast to zero and a jammed traffic is observed. 
For high densities  discontinuous nature  of car velocities is manifested, when
peaks at multiples of five appear in the diagram. For $X=5$ $(\rho =1) $  
practically all the cars have velocity equal to zero.  

The system of cars may exist in two phases:  a fast, free-flow phase and slow, 
jammed phase. Nevertheless, contrary to the standard models of statistical 
mechanics, they cannot coexist as  a collision between fast and slow group 
would take place.  
Thus, the 
transition from a local maximum to absolute maximum of probability is very 
improbable, and a strong hysteresis occurs in the system, observed also 
experimentally [12].  
 
If the reaction time of the drivers is nonzero, the velocity of cars decreases 
with group length faster, the most probable velocity decreases continuously,  
and no first order phase transition takes place. The transition between 
free-flow and jammed phase is continuous and probability of velocities, i.e., 
entropy as a function of velocities is concave.
 This case for reaction time $\tau=0.5$ is shown in Fig.~2.

When the car velocity spread is large, $p_2/p_0=0.045$, most of the cars are 
in the 
jammed phase, as depicted in Fig.~3. Only a small peak of a free-flow traffic 
can be seen in the Figure.

For low braking abilities of cars, $a=0.5$, no free flow, even for density 
$\rho=0.033$,  low velocity spread $p_2/p_0=0.005$, and zero reaction time,  
 exists, as shown in Fig.~4. At high density, there is convenient for them 
not to brake at all, so that they form platoons in which all the cars have the 
same velocity -- in our case an integer one. This is the reason why in Fig.~4 
the most probable total velocities up to $V=40$  are multiples of five.

In the microcanonical approach the subsystem and the reservoir have the same 
density. Density fluctuations are described in the canonical approach where 
the 
length of the subsystem changes while the total length of the system remains 
constant. The probability that the total velocity of the subsystem is $V$ and 
its length is $X$ in the system with mean density is $(N+N_{\rm r})/  L_{\rm 
s}$ is proportional  to $W(V,X)$. It is depicted in Fig.~5 
for  $a=4$, $p_2/p_0=0.025$, $\tau=0$ and mean length $\langle X\rangle =50$. 
Low and high density 
fluctuation are suppressed due to small number of states of the reservoir and 
subsystem, respectively. 
 
For inefficient car brakes the platoon structure of is seen also in SPS of 
a canonical ensemble, but only for high density fluctuations. At low density 
the probability maxima disappear and the cars may easily change the velocity 
of the platoon.  It is shown in Fig.~6 for $a=0.5$.

Our results are in qualitative agreement with those obtained from the 
approach of Nagel and Schreckenberg  [4]. As it may be seen from Fig.~1, for 
low densities the velocity of the cars  with increasing density remains 
constant, i.e., the flow linearly increases. At the density where the confined 
regime 
starts, the velocity and flow sharply decreases, and at $\rho = 1$ the flow is 
practically equal to zero. We have got not only the mean velocity in the system
but also  distributions of velocities of small groups of cars for each density 
and probabilities of density fluctuations for various velocities.   
The distribution is widest near the phase transition point, where very high 
and very low velocities are equally probable.

A statistical approach to traffic flow was also chosen by Mahnke et al. [5]    
who, starting from grandcanonical ensemble,    introduced 
thermodynamical potential of standard Hamiltonin statistical mechanics, 
potential energy of interactions between vehicles, etc. We have developed the 
statistics of the system in analogy with developing statistical mechanics, 
starting from a microcanonical ensemble, but now  not fixing the energy but 
other 
quantities characterizing the system.  The conditional probabilities of car 
velocities for given headway and velocity of the car ahead were assumed to be 
known, and probabilities of states differing in other quantities were taken  
equally probable, satisfying the principle of maximum entropy.

Number of cars on the road is small, much smaller than number of particles in 
most physical systems studied in statistical mechanics. Therefore it represent 
a suitable area for application of ideas recently developed for very small 
physical systems [6--10].

\bigskip\bigskip
\ni{\bfb Acknowledgement}

\bigskip
\ni We acknowledge support from VEGA grant No. 2/6071/2006.

\bigskip\bigskip
\ni{\bfb References}

\bigskip

\ninepoint
\item{[1]} 
D. E. Wolf, M. Schreckenberg, and A. Bachem (Eds.), Traffic and Granular 
Flow, World Scientific, (1996);
D. Helbing, H.J. Herrmann, M. Schreckenberg, D.E. Wolf (Eds.),
Traffic and GranularFlow  99, 
Springer, Berlin, (2000); 
M. Fukui, Y. Sugiyama, M. Schreckenberg, D.E. Wolf (Eds.),
Traffic and GranularFlow  01, 
Springer, Heidelberg, 
(2003); 
S.P. Hoogendoorn, P.H.L. Bovy, M. Schreckenberg, D.E. Wolf (Eds.),
Springer, Heidelberg, (2005).

\item{[2]} 
 D. Helbing, Rev. Mod. Phys., 73, 1067 (2001).

\item{[3]} 
 R. Herman, K. Gardels, Sci. Am. 209, 35 (1963)

\item{[4]} 
 K. Nagel and M. Schreckenberg, J. Phys. I (France) {2}, 2221 (1992).  

\item{[5]} 
R. Mahnke, J. Hinkel, J. Kaupu\v zs, and H. Weber,
Thermodynamics of traffic flow,
cond-mat/0606509 (2006).

\item{[6]} 
D. H. E. Gross, Microcanonical thermodynamics: Phase
transitions in ``small'' systems, Lecture Notes in Physics,
66, World Scientific, Singapore, (2001).

\item{[7]} 
R. J. Creswick, Physical Review E,  52,  5735 (1995) 

\item{[8]} 
D. H.~E. Gross,
Geometric Foundation of Thermo-Statistics,
Phase Transitions, Second Law of Thermodynamics,
but without Thermodynamic Limit,
cond-mat/0201235 (2002)  

\item{[9]} 
M. Kastner, M. Promberger, and A. H\"uller, J. Stat. Phys.
99, 1251 (2000).

\item{[10]} 
H. Behringer, On the structure of the entropy surface of microcanonical 
systems, Mensch und
Buch Verlag, Berlin, (2004).

\item{[11]} 
R. Jiang, M.B. Hu, B. Jia, R.L. Wang, and Q.S. Wu,
Eur. Phys. J. B 54, 267 (2006).

\item{[12]} 
D. Chowdhury, L. Santen and A. Schadschneider, Physics
Reports 329, 199 (2000).

\bye

%% file: A4M.tex
\magnification=\magstep1
\hsize=16.5truecm
\vsize=25truecm
\baselineskip 0.55truecm plus 0.01truecm minus 0.01truecm
\parindent=15pt
\hoffset=-1.5cm

\font\rmb=cmr10 scaled \magstep1
\font\bfb=cmbx12

\font\bfbb=cmbx10 scaled \magstep2

\font\ninerm=cmr9  \font\sixrm=cmr6
\font\ninei=cmmi9  \font\sixi=cmmi6
\font\ninesy=cmsy9 \font\sixsy=cmsy6
\font\ninebf=cmbx9 \font\sixbf=cmbx6
\font\nineit=cmti9 
\font\ninesl=cmsl9

\def\ninepoint{\def\rm{\fam0\ninerm}%
\textfont0=\ninerm\scriptfont0=\sixrm\scriptscriptfont0=\fiverm%
\textfont1=\ninei\scriptfont1=\sixi\scriptscriptfont1=\fivei%
\textfont2=\ninesy\scriptfont2=\sixsy\scriptscriptfont2=\fivesy%
\textfont3=\tenex\scriptfont3=\tenex\scriptscriptfont3=\tenex%
\textfont\itfam=\nineit\def\it{\fam\itfam\nineit}%
\textfont\slfam=\ninesl\def\sl{\fam\slfam\ninesl}%
\textfont\bffam=\ninebf\scriptfont\bffam=\sixbf%
\scriptscriptfont\bffam=\fivebf\def\bf{\fam\bffam\ninebf}%
\normalbaselineskip=0.37cm%
\setbox\strutbox=\hbox{\vrule height8pt depth3pt width0pt}%
\let\sc=\sevenrm\normalbaselines\rm}

\let\ltlt=\ll

\def\ll{\leftline}
\def\cl{\centerline}

\def\ni{\noindent}
\def\({\left(}
\def\){\right)}
\def\[{\left[}
\def\]{\right]}

\def\C{\ifmmode \kern1pt^\circ\kern-1.5pt\hbox{C}\else 
$\kern1pt^\circ\kern-1.5pt${C}\fi}

\let\eqln=\eqalignno
\def\dst{\displaystyle}

\def\bitem#1{\lastdef\bgroup\setbox99=\hbox{#1\enspace}\dimen98=\parindent 
\parindent= \wd99 \setbox97=\hbox{#1}\parskip0pt}
\def\eitem{\par\egroup}
\def\itik#1{\item{\hbox to \wd97{#1\hfil}}}

\def\bcentr{\bgroup\parindent0pt\parfillskip0pt\leftskip0pt plus1fil
\rightskip0pt plus1fil}
\def\ecentr{\par\egroup}

\newdimen\rulewidth
\rulewidth=0.4pt

\def\Psboxrot#1#2{%
\def\epsfsize##1##2{#1##1}%
\setbox0=\vbox{\epsfbox{#2.eps}}
\dimen1=\wd0
\vbox{\hrule 
height\rulewidth\hbox{\vrule width\rulewidth 
\vbox to\wd0{\vss\hbox to\ht0{\raise\dimen1\pstransform{90 
rotate}{\box0}\hss}}\vrule width\rulewidth}\hrule height\rulewidth}}

\def\Ps#1#2#3{
\def\epsfsize##1##2{#2##1}
\vskip2mm
\cl{%
\vbox{\null\hrule height\rulewidth\hbox{\vrule 
width\rulewidth \epsfbox{#1.eps}%
\vrule width\rulewidth}\hrule height\rulewidth}}
\vskip5mm
{\ninepoint\parindent0pt\pcn
#3
\bigskip} 
\endinsert
}

\def\Psn#1#2#3{
\def\epsfsize##1##2{#2##1}
\vskip2mm
\cl{%
\vbox{\null\hrule height\rulewidth\hbox{\vrule 
width\rulewidth \epsfbox{#1.eps}%
\vrule width\rulewidth}\hrule height\rulewidth}}
\vskip5mm
{\ninepoint\parindent0pt\pcn
#3
\bigskip}}

\def\pcn{\leftskip=0pt plus1fil \rightskip=0pt plus-1fil
\parfillskip=0pt plus2fil}

\input epsf